\documentclass[5p,times]{elsarticle}
\usepackage{lineno,hyperref}
\modulolinenumbers[5]
\usepackage[utf8]{inputenc} 
\usepackage[T1]{fontenc}    
\usepackage{hyperref}       
\usepackage{url}            
\usepackage{booktabs}       
\usepackage{amsfonts}       
\usepackage{nicefrac}       
\usepackage{microtype}      
\usepackage{graphicx}
\usepackage{doi}
\usepackage{hyperref} 
\usepackage{amsmath}
\usepackage{upgreek}
\usepackage{color}
\usepackage{soul}
\usepackage{algorithm2e}
\usepackage{algpseudocode}
\usepackage{graphicx}
\usepackage{lscape}
\usepackage{tabularx}
\usepackage{multirow}
\usepackage{caption} 
\journal{Journal of Information Security and Applications}

\begin{document}
	
	\begin{frontmatter}
		
		\title{Review on the Feasibility of Adversarial Evasion Attacks and Defenses \\ for Network Intrusion Detection Systems}

		\author[mymainaddress,mysecondaryaddress]{Islam Debicha\corref{mycorrespondingauthor}}
		\cortext[mycorrespondingauthor]{Corresponding author}
		\ead{islam.debicha@ulb.be}
		
		\author[mymainaddress]{Benjamin Cochez\corref{mycorrespondingauthor}}
		\ead{benjamin.cochez@ulb.be}
		\author[mythirdaddress]{Tayeb Kenaza}
		\author[mysecondaryaddress]{Thibault Debatty}
		\author[mymainaddress]{Jean-Michel Dricot}
		\author[mysecondaryaddress]{Wim Mees} 
		
		\address[mymainaddress]{ Cybersecurity Research Center, Université Libre de Bruxelles, 1000 Brussels, Belgium}
		\address[mysecondaryaddress]{Cyber Defence Lab, Royal Military Academy, 1000 Brussels, Belgium}
		\address[mythirdaddress]{Computer Security Laboratory, Ecole Militaire Polytechnique, Algiers, Algeria}

		\begin{abstract}

			Nowadays, numerous applications incorporate machine learning (ML) algorithms due to their prominent achievements. However, many studies in the field of computer vision have shown that ML can be fooled by intentionally crafted instances, called adversarial examples. These adversarial examples take advantage of the intrinsic vulnerability of ML models. Recent research raises many concerns in the cybersecurity field. An increasing number of researchers are studying the feasibility of such attacks on security systems based on ML algorithms, such as Intrusion Detection Systems (IDS). The feasibility of such adversarial attacks would be influenced by various domain-specific constraints. This can potentially increase the difficulty of crafting adversarial examples. Despite the considerable amount of research that has been done in this area, much of it focuses on showing that it is possible to fool a model using features extracted from the raw data but does not address the practical side, i.e., the reverse transformation from theory to practice. For this reason, we propose a review browsing through various important papers to provide a comprehensive analysis. Our analysis highlights some challenges that have not been addressed in the reviewed papers. 
			
		\end{abstract}
		
		
	\end{frontmatter}
	
	
	\section{Introduction}
	With the development of new technologies and the increasing evolution of Internet interconnections, security is now a crucial issue. To defend against the various existing attacks, some defensive systems use ML algorithms such as anomaly-based IDS (AIDS) which is currently a widely used security tool \cite{gomez2009design} due to its ability, among other benefits, to detect unknown attacks, i.e., zero-day  attacks\cite{garcia2009anomaly,khraisat2019survey}. Nevertheless, recent studies have shown that ML models in general, and deep neural networks in particular, are vulnerable to so-called adversarial attacks and that the addition of small specifically designed perturbations can mislead the classifier \cite{he2020towards,bae2018security,ilyas2019adversarial}.
	
	Today, problems related to the security of ML-based IDS are an active research topic \cite{debicha2022tad,zhang2022adversarial}. In the context of network-based IDS, this means that it is possible to design specific perturbations to be added to network traffic by manipulating certain properties, such as packet size, or send/receive time and duration. These perturbations can mislead a classifier into identifying attack traffic as benign and thus evading the intrusion detection system.

	In addition, a significant amount of research on the impact of adversarial learning in computer vision has been transferred into intrusion detection. Initial results have shown that the classifiers used in AIDS are also vulnerable to these algorithms. A typical approach used by researchers is to focus on the theoretical aspect of the problem by setting simplifying assumptions and focusing only on the feature space \cite{wang2018deep,yang2018adversarial,martins2019analyzing}. However, unlike computer vision where the created perturbations have relatively few constraints, a valid network traffic perturbation must satisfy many domain-specific constraints (both semantic and syntactic). These domain-specific constraints ensure that the added perturbation will generate valid network traffic enabling the transition from feature space to traffic space. Unfortunately, network-specific constraints are often not considered or only to a limited extent. This means that the feasibility of attacks from a realistic point of view is not fully considered. Some researchers \cite{han2021evaluating,sadeghzadeh2021adversarial,chen2020generating} have decided to take a different approach to deal with the problem by limiting the need for feature knowledge by directly manipulating the traffic space.

	Our main contribution is therefore a revised review of the state of the art providing a new aspect based on the feasibility of attacks. We also provide an update on new contributions that have been produced recently concerning the feasibility of attacks in real settings:
	\begin{itemize}
		\item We propose a complete analysis, for each selected paper, on the real feasibility of the proposed attacks by demonstrating whether or not the constraints of the domain are respected. 
		\item We propose an analysis of the defenses used in the papers studied to highlight the strengths and weaknesses of each. 
		\item We identify some realistic aspects that should be considered for future studies of the impact of adversarial attacks on IDSs.
	\end{itemize}
	We believe this review will help future research in creating realistic attacks that consider the full context of the IDS domain. We also believe that it will help in the understanding and creation of new defense mechanisms that improve the robustness of the developed models. 
	
	The rest of the paper is structured as follows. Section \ref{sec:background} gives a theoretical reminder introducing the key concepts used in the reviewed papers. Section \ref{sec:relwork} summarizes previous reviews that have been conducted on the subject. Section \ref{sec:attacks} describes the most commonly used state-of-the-art attacks in the literature. Section \ref{sec:defenses} shows the most popular defense mechanisms used in the literature to counter the attacks described in Section \ref{sec:attacks} . Section \ref{sec:studiedpaper} contains a detailed analysis of the realism of the selected papers. Section \ref{sec:discuss} discusses the actual feasibility of the attacks present in Section \ref{sec:studiedpaper} and the challenges associated with them. Section \ref{sec:conclu} concludes the paper by providing the key points that have been discussed.
	
	\section{ Background and related work}
	
	\label{sec:background}
	
	\subsection{Anomaly-based IDS}
	The increasing development of new threats targeting network infrastructures worldwide has pushed researchers to develop new defenses. Due to the huge number of undiscovered attacks, most defense mechanisms are unable to cope with such threats. To mitigate this problem, solutions such as Anomaly-Based IDS ( AIDS ), which can detect some of these previously unknown attacks through the use of statistics and machine learning algorithms, are gaining popularity. AIDS have many properties, among which we note their ability to be deployed in a network (NIDS) or directly on a host (HIDS). As far as NIDS are concerned, two different types can be found: packet-based and flow-based. The data used by NIDS to collect this information can come from different sources such as network protocols like NetFlow/IPFIX, SNMP, or directly from an agent. NIDS can also use application logs from anti-virus or firewalls. To measure the performance of NIDS, different metrics are used such as true positive rate (TPR), true negative rate (TNR), false positive rate (FPR), false negative rate (FNR), accuracy, precision, F1 score, error rate, area under the curve (AUC). All these metrics are derived from the \textit{confusion matrix} shown in \autoref{tab:idsmetrics}.
	\begin{table}[h]
		\centering
		\begin{tabular}{lll}
			\multicolumn{3}{c}{Predicted Class}    \\
			\cmidrule(r){2-3}
			Actual Class & Anomaly & Normal \\
			\midrule
			Anomaly & True Positive (TP)  & False Negative (FN)     \\
			Normal  & False Positive (FP) & True Negative (TN)      \\
			\bottomrule
		\end{tabular}
		\vspace{0.2cm}
		\caption{IDS confusion matrix}
		\label{tab:idsmetrics}
	\end{table}

	\subsection{Threat model}
	Although there are several types of threats, an attacker often seeks to violate one of the following properties: confidentiality, integrity, authenticity, and availability.
	
	In terms of threat modeling, there are two important points to consider, namely the knowledge restriction corresponding to the complexity of the attack and the objective of the attack, corresponding to the capability of the attack.
	
\begin{figure*}
		\centering
		\includegraphics[width=1.0\linewidth]{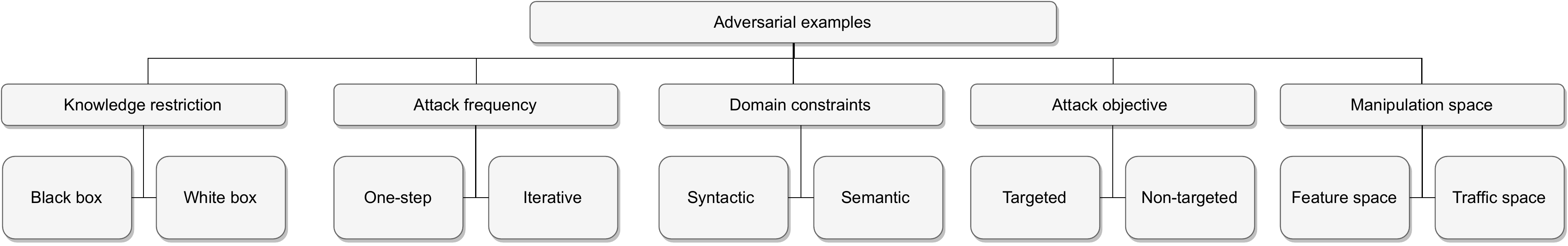}
		\caption{Classification of adversarial examples}
		\label{fig:classification}
	\end{figure*}

	\paragraph{Knowledge restriction}
	As shown in Figure \ref{fig:classification}, attacks can be conducted in two forms, black box or white box. The white box attack means that the adversary knows everything about the training dataset and the model architecture, in particular all the parameters and meta-parameters which are for example the inputs, the gradients (for DNNs), the tree depth (for decision trees) or the number of neighbors (for K-nearest neighbors) as well as the chosen cost function or the type of optimizer (e.g., ADAM or RMSProp) in case of neural networks.  The black box refers to the fact that the attacker knows nothing about the target model, i.e., the architecture of the model and the dataset used. The attacker can only send requests to the targeted model and receive answers in the form of decisions or probabilities (logits). He must, therefore, without knowing any information about the model, approximate a decision boundary similar to that of the target model to be able to craft adversarial samples. Another option for black box attacks is exploiting transferability.  An attacker can create a surrogate model, similar in functionality to the targeted model, craft adversarial instances to fool the surrogate model, and then transfer those instances to the targeted model so that it will also be fooled. Black-box attacks are more complicated to perform since less knowledge is available, but also because more computational resources are needed to accommodate this accumulated knowledge (queries). 
	
	\paragraph{Attack objective}
	Another relevant property of an attack is its objective. There are two different types of objectives, the untargeted attack, and the targeted attack. A non-targeted attack is easier to perform since all the attacker has to do is trick the machine learning model without any particular considerations. Two possible scenarios can be expected. The first is confidence reduction, which means that the attacker simply wants to decrease the performance of the model while maintaining the overall functionality. The second scenario is misclassification. In this case, the adversary's goal is to trick the model into misclassifying without specific constraints. In a targeted attack, the adversary's goal is to force an ML model to produce the desired output by manipulating the input. This type of objective is therefore more complicated to achieve because it requires manipulating the model in a specific direction, unlike a non-targeted attack that is not limited to a certain target. There are two variants of targeted attacks that can be highlighted. The first is targeted misclassification, which means that an attacker wants to cause misclassification in a certain target class with any input. The other variant is source/target misclassification, which means that an attacker wants to cause misclassification in a certain target class with a certain input. This particular goal is the most difficult to achieve.
	
	\subsection{Adversarial examples}
	Adversarial learning refers to the problem of designing attacks against machine learning as well as defenses against these attacks. Depending on the phase in which the attack is carried out, adversarial attacks can be divided into poisoning and evasion attacks. This paper focuses on evasion attacks. This choice is due to the fact that this review wants to focus on the most realistic aspects of adversarial attacks against NIDS. The problem with poisoning attacks is that they require the ability to directly manipulate the model training data. It is clear that in a realistic scenario, the attacker's knowledge will be limited, and it will be less possible to manipulate the model before its training phase.
	
	The creation of adversarial examples can be expressed as an initial problem formulation as defined in Eq. \ref{eq:advex}.
	\begin{equation}
		\label{eq:advex}
		\begin{split}
			\text{Minimize: } D(x, x + \delta)\\
			\text{Such that: } C(x + \delta) = t \text{---- constraint 1}\\
			x + \delta \in [0,1]^n \text{---- constraint 2} 
		\end{split}
	\end{equation}
	where we want to minimize the distance between the original element and the adversarial element D(x, x + $\delta$) respecting 2 constraints. The first is that the classification C of x + $\delta$ must be classified as the target label t. The second is that x + $\delta$ must be a valid element.
	
	According to the work of Szegedy \textit{et al. }\cite{szegedy2013intriguing}, adversarial examples exploit the fact that neural networks have "blind spots". The cause of this "blind spot" effect would be due to the models being non-linear and trying to behave linearly as concluded by Goodfellow \textit{et al. }\cite{goodfellow2014explaining}. These adversarial examples have certain properties described below. 
	
	\paragraph{L\textsubscript{P} norms}
	To compute the distance between the original element $x$ and the perturbed element $x_{adv}$, an L\textsubscript{P} norm (i.e., distance metric) is used such as $L_0$,$L_1$, $L_2$ and $L_\infty$ allowing to define the boundary of adversarial examples. These norms are thus used to minimize the perturbation rate used to generate the adversarial example. The most common norms used by adversarial algorithms are:
	
	$L_0$: This distance metric counts the number of features of $x$ modified in $x_{adv}$. This metric only takes into account the number of modified features regardless of the perturbation rate introduced in each feature.
	
	$L_1$: This norm represents the Manhattan distance between $x$ and $x_{adv}$ as defined in Eq. \ref{eq:manathan}.
	\begin{equation}
		\label{eq:manathan}
		L_1 = |x_1 - {x_1}_{adv}| +  ... + |x_n - {x_n}_{adv}|
	\end{equation}
	
	$L_2$: This norm calculates the Euclidean distance or the mean-squared error between $x$ and $x_{adv}$ as shown in Eq. \ref{eq:euclid}.
	\begin{equation}
		\label{eq:euclid}
		L_2 = \sqrt{(x_1 - {x_1}_{adv})^2  + ... + (x_n - {x_n}_{adv})^2}
	\end{equation}
	
	$L_\infty$: This norm gives the largest change among all features of $x_{adv}$ compared to $x$ and it's defined in the following Eq. \ref{eq:infin}.
	\begin{equation}
		\label{eq:infin}
		L_\infty = max(|x_1 - {x_1}_{adv}|, ..., |x_n - {x_n}_{adv}|)
	\end{equation}

	\paragraph{Attack frequency}
	Attack frequency is a property that defines whether the attack is executed in a one-step iteration or requires several. Thus, there are two types of attacks: one-step attacks and iterative attacks. One-step attacks mean that the adversarial examples are generated by an algorithm that executes only once, i.e., it does not iterate multiple times to optimize the adversarial example. Thus, one-step attacks are faster but less optimized. Iterative attacks on the other hand use iterative functions to generate adversarial examples so that it maximizes their efficiency but takes more time.
	
	\paragraph{Domain constraints}
	The feasibility of adversarial attacks is domain-specific and is influenced by several constraints. These constraints can be divided into two main categories: syntactic constraints and semantic constraints. The following syntactic constraints were originally discussed by Merzouk \textit{et al. }\cite{merzouk2022investigating}. As for the semantic links, the exact definition of these constraints is difficult since they are specific to each domain and even to each type of feature used. However, we draw on the work of Hashemi \textit{et al. }\cite{hashemi2019towards}, and Teuffeunbach \textit{et al. }\cite{teuffenbach2020subverting} in the IDS domain to provide a generalization of three different groups with different semantic links.
	
	\textit{Syntactic constraints} concern all those related to syntax, e.g., out-of-range values, non-binary values and multiple category membership. Out-of-range values are values that exceed a theoretical maximum value that cannot be exceeded, for example, a float between 0 and 1 or an integer between 0 and 255. Non-binary values are entries that violate the binary nature of a feature and multiple category membership are values that violate the one-hot encoding concept.
	
	\textit{Semantic Links} represent the links that certain features may have with each other. These features can be grouped into three distinct groups, each with different semantic properties. The first set includes features that cannot be modified (e.g., IP address, protocol type). The second group includes features that can be directly modified (number of forward packets, size of the forward packet, flow duration, ...). The last group concerns the features that depend on the second group. They must be recalculated based on the latter (number of packets/second or average forward packet size).

	This implies that the complexity of generating realistic adversarial examples varies with the different types of data used to represent the domain, such as numerical (continuous or discrete) or categorical data. It also depends on the context in which the model is located (such as network traffic). The NIDS domain is therefore strongly affected by both semantic and syntactic constraints as it uses heterogeneous data types, and its context requires several semantic links most of the time unlike other domains such as computer vision.
	
	\paragraph{Manipulation space}
	An essential property of a realistic adversarial instance is the ability of an attacker to modify its characteristics. In theory, it is possible to directly modify the features of adversarial instances. However, in real-world scenarios, this approach is considered unsuitable for certain domains such as IDSs that analyze network traffic. This is mainly due to the fact that the feature extraction process (i.e., from raw traffic to feature space) is not a fully reversible process, unlike other domains such as computer vision. This means that features can be extracted, and modified but not easily reintroduced into network traffic due to the semantic links between features. Moreover, direct feature modification requires full knowledge of the feature extraction process used by the IDS in order to respect the syntactic or semantic constraints assigned to them. We can therefore deduce that working on the feature space is not very realistic. For this reason, recent studies \cite{han2021evaluating,sadeghzadeh2021adversarial,chen2020generating,hashemi2019towards} propose to manipulate directly the raw network traffic so that it is not necessary to know the features used, nor to transform the feature values into traffic form. In this way, we can distinguish two manipulation spaces, the feature-based and the traffic-based.
	
	\subsection{Related work}
	\label{sec:relwork}
	Numerous research studies on the real impact of adversarial attacks have already been extensively conducted in the compute vision field, which has also urged researchers to study the issue in the cybersecurity field. Today, the number of papers on this topic is rapidly increasing and the actual impact of these attacks in a real-world scenario seems to be getting clearer. To help the community gain more insight into the topic, we analyze the important aspects of the feasibility of adversarial attacks by comparing the different research and reviews on the topic, especially those related to IDS.

	In the review proposed by Reza \textit{et al. }\cite{wiyatno2019adversarial}, the authors focus on giving a better understanding of adversarial examples in the computer vision domain. They propose an analysis of numerous attacks and defenses dedicated to this domain. Among these attacks, some are more realistic as they are directly applicable to a real-world scenario. However, this review does not provide any information about the implication of these attacks and defenses in the IDS domain. In addition, the review does not address the topic of domain constraints, nor the attack manipulation space. 
	
	Vitorino \textit{et al. }\cite{vitorino2022adaptative} took an interesting approach in their paper to analyze, from the point of view of domain constraints, the suitability of adversarial attacks for the IDS domain. They showed that most of the state-of-the-art attacks, initially dedicated to computer vision, were not suitable for the IDS domain as they did not comply with these constraints. However, the paper does not address the manipulation space used, nor the problems related to the respect of semantic and syntactic constraints found in papers dealing with attacks against IDSs. In addition, defenses against adversarial examples in the IDS domain are also not addressed. 
	
	The study proposed by McCarthy \textit{et al. }\cite{mccarthy2022functionality} proposes the analysis of several attacks and defenses in different domains of cybersecurity, namely intrusion detection, malware detection, and anomaly detection in industrial systems. They pointed out some constraints related to adversarial algorithms. Our review further elaborates on the manipulation space property, as well as a discussion of semantic and syntactic constraints that are not discussed in detail in their paper. In addition, our work surveys more recent papers.
	
	The paper by Apruzzese \textit{et al. }\cite{apruzzese2021modeling} provides interesting insights into the manipulation space used in defining attacks as problem or feature-based. This work also provides an in-depth analysis of the different learning phases of the model by articulating the feasibility at all levels of the machine-learning pipeline. Our work differs by providing an analysis of more recent work on the topic and an explanation of each paper based on the domain constraints analysis. In addition, their work does not include an overview of possible defenses. 
	
	Martins \textit{et al. }\cite{martins2020adversarial} provides a comprehensive overview of adversarial attacks against IDS and malware classifiers. They also describe the state of the art of defenses. However, this review does not include a discussion of the feasibility aspect of adversarial attacks and defenses. In our contribution, an analysis of the realistic aspect of the state-of-the-art defenses and attacks is introduced with an explanation of their feasibility.
	
	\section{Adversarial strategies}
	\label{sec:attacks}
	In this section, we present state-of-the-art adversarial attacks, classified into white-box and black-box algorithms. A list of these attacks can be found in \autoref{tab:attacks}.

	\begin{table*}[]
		\centering
		\resizebox{\textwidth}{!}{%
			\def\arraystretch{1.5}
			\begin{tabular}{|c|c|c|c|c|c|c|c|}
				\hline
				\textbf{\begin{tabular}[c]{@{}c@{}}Knowledge\\ restriction\end{tabular}} &
				\textbf{Attack} &
				\textbf{Advantages} &
				\textbf{Disadvantages} &
				\textbf{Norm} &
				\textbf{\begin{tabular}[c]{@{}c@{}}Attack \\ frequency\end{tabular}} &
				\textbf{\begin{tabular}[c]{@{}c@{}}Attack \\ objective\end{tabular}} &
				\textbf{\begin{tabular}[c]{@{}c@{}}Suitable\\  for IDS\end{tabular}} \\ \hline
				\multirow{16}{*}{White box} &
				L-BFGS &
				Efficient in generating adversarial instances &
				highly computationally demanding &
				$L_2$ &
				I &
				T &
				/ \\ \cline{2-8} 
				&
				FGSM &
				Calculation time efficiency &
				\begin{tabular}[c]{@{}c@{}}All features, including non-modifiable ones, \\ are perturbed.\end{tabular} &
				\begin{tabular}[c]{@{}c@{}}$L_2$\\ $L_\infty$\end{tabular} &
				O &
				T \& NT &
				/ \\ \cline{2-8} 
				&
				PGD/BIM &
				\begin{tabular}[c]{@{}c@{}}More efficient than FGSM \\ due to its iterative nature\end{tabular}&
				\begin{tabular}[c]{@{}c@{}}When compared to FGSM, it is  \\ more computationally expensive\end{tabular}&
				\begin{tabular}[c]{@{}c@{}}$L_2$\\ $L_\infty$\end{tabular} &
				I &
				T \& NT &
				/ \\ \cline{2-8} 
				&
				
				JSMA &
				\begin{tabular}[c]{@{}c@{}}Successfully deceive a model by  \\ changing a few input features\end{tabular} &
				More computationally demanding compared to FGSM &
				$L_0$ &
				I &
				T &
				Could \\ \cline{2-8} 
				&
				DeepFool &
				Produce significantly smaller perturbations than FGSM  &
				\begin{tabular}[c]{@{}c@{}}Perturbations are sub-optimal \\ More time consuming than FGSM and JSMA\end{tabular} &
				$L_2$ &
				I &
				NT &
				/ \\ \cline{2-8} 
				&
				C\&W &
				\begin{tabular}[c]{@{}c@{}}Empirically shown to be more effective than other attacks\\ Successfully circumvented many adversarial defenses\end{tabular} &
				More computationally intensive than previous attacks &
				\begin{tabular}[c]{@{}c@{}}$L_0$\\ $L_2$\\ $L_\infty$\end{tabular} &
				I &
				T \& NT &
				/ \\ \cline{2-8} 
				&
				EAD &
				\begin{tabular}[c]{@{}c@{}}Produce highly transferable instances\\ Successfully bypassed defensive distillation\end{tabular} &
				More time-consuming than other attacks such as FGSM &
				$L_1$ &
				I &
				T \& NT &
				/ \\ \hline
				\multirow{9}{*}{Black box} &
				Substitute Model &
				\begin{tabular}[c]{@{}c@{}}Successfully defeats gradient masking-based defenses \\Feasible against non-differentiable	models\end{tabular} &
				Not as effective as white box attacks &
				/ &
				I &
				T \& NT &
				yes \\ \cline{2-8} 
				&
				ZOO &
				Its performance is comparable to that of C\&W &
				\begin{tabular}[c]{@{}c@{}}Empirically slower than Substitute Model attack \\ Needs a significant amount of queries to the target classifier\end{tabular} &
				$L_2$ &
				I &
				T \& NT &
				/ \\ \cline{2-8} 
				&
				Boundary &
				\begin{tabular}[c]{@{}c@{}}Knowledge of the victim's model is not required\\ Deliver comparable results to white box attacks\end{tabular} &
				\begin{tabular}[c]{@{}c@{}}Require a substantial number of queries \\ to find high quality adversarial examples \end{tabular} &				
				
				$L_2$ &
				I &
				T \& NT &
				/ \\ \cline{2-8} 
				&
				OPT &
				Requires fewer queries compared to ZOO and Boundary &
				Being a query-based attack, it can be easily detected &
				\begin{tabular}[c]{@{}c@{}}$L_2$\\ $L_\infty$\end{tabular} &
				I &
				T \& NT &
				/ \\ \cline{2-8} 
				&
				
				GAN/WGAN &
				Can create samples that differ from those used in training &
				This attack can be computationally heavy and highly unstable &
				/ &
				I &
				T \& NT &
				yes \\ \hline
			\end{tabular}
		}
		\caption{List of well-known state-of-the-art evasion attacks (I: iterative, O: one-time, T: targeted, NT: non-targeted)}
		\label{tab:attacks}
	\end{table*}

	\subsection{White-Box algorithms}
	\paragraph{Limited-memory Broyden Fletcher Goldfarb Shanno (L-BFGS)}
	
	The idea of this iterative attack is to produce an instance $x_{adv}$ similar to the initial instance $x$ under the distance $L_2$ but have $x_{adv}$ classified as another target class using the L-BFGS box constraint. For this, Szegedy \textit{et al. }\cite{szegedy2013intriguing} explain that it's possible to express the initial problem as a constrained minimization problem to generate targeted adversarial examples as illustrated in Eq. \ref{eq:lbfgseq}. 
	\begin{equation}
		\label{eq:lbfgseq}
		\begin{split}
			\text{Minimize: }||x - x_{adv}||_2 \\
			\text{Such that: }C(x_{adv}) = t \\
			x_{adv} \in [0,1]^n
		\end{split}
	\end{equation}
	Since this problem is difficult to solve, they adapt it into an easier-to-handle variant, as shown in Eq. \ref{eq:lbfgseq2}.
	\begin{equation}
		\label{eq:lbfgseq2}
		\begin{split}
			\text{Minimize: }c.|x - x_{adv}| + J(x_{adv}, t)\\
			\text{Such that: }x_{adv} \in [0,1]^n
		\end{split}
	\end{equation}
	where x is the input element, $x_{adv}$ is the adversarial element, c is a positive constant, J is the loss function and t is the target label. 
	
	On the other hand, while L-BFGS is an effective attack, it can be time-consuming due to the use of the linear search method to find an optimal c. 
	
	\paragraph{Fast gradient sign method (FGSM)}
	This one-step algorithm was developed by Goodfellow \textit{et al. } in their 2014 paper \cite{goodfellow2014explaining}. The idea of FGSM is to generate perturbation using gradient ascent to maximize the loss function. FGSM can be used as a targeted or untargeted attack and originally runs under the $L_\infty$ norm but is easily adaptable for the $L_2$ norm. FGSM is a very fast algorithm for generating adversarial instances even if the adversarial samples are not optimized because it does not minimize the generated perturbation. This algorithm is very efficient, in most cases, at creating adversarial perturbations in a time-efficient manner. It can be defined by the following Eq. \ref{eq:fgsmeq}. 
	\begin{equation}
		\label{eq:fgsmeq}
		x_{adv} = x + \epsilon * sign(\nabla_x J(x, y))
	\end{equation}
	where $\epsilon$ is the variable allowing control of the amount of perturbation, y is the desired label, and the input x. The main disadvantage of this attack from a network traffic perspective is that all features are modified, making it less practical in real-life scenarios. 
	
	\paragraph{Basic Iterative Method/Projected Gradient Descent (BIM/PGD)}
	This is an improvement of FGSM where the algorithm iteratively increases the amount of perturbation to cause misclassification. It is more efficient than the classical FGSM in terms of misclassification, but on the other hand, this attack takes more time to create adversarial examples. PGD is an algorithm proposed by Aleksander Madry \textit{et al. }\cite{madry2017towards} and BIM is proposed by Alexey Kurakin \textit{et al. }\cite{kurakin2018adversarial}. Both attacks are quite similar as they use, at each iteration, a projection function to project the adversarial examples into the $\epsilon-ball$ which can be $L_2$ or $L_\infty$, as shown in Eq. \ref{eq:pgdeq}.
	
	\begin{equation}
		\label{eq:pgdeq}
		x^t_{adv} = Proj[x^{t-1} + \epsilon * sign(\nabla_x J(x^{t-1}, y))]
	\end{equation}
	where ${x_0}_{adv} = 0$ and $Proj$ is the projection function.
	
	The main difference between the BIM and PGD versions of the attack concerns the initialization of the attack. Indeed, BIM sets the value of the original point as the initialization point while PGD starts the attack at a random point using the L$_\infty$ norm. Moreover, at each restart, a new random point is chosen. Since the results of these two attacks are generally quite similar, it is common to use only one of them when testing.
	
	\paragraph{DeepFool}
	This attack proposed by Moosavi-Dezfooli \textit{et al. }\cite{moosavi2016deepfool} works as an untargeted attack and iteratively generates small perturbations to fool the classification. This algorithm uses the $L_2$ norm to generate these perturbations. To do so, this attack determines the nearest hyperplane for an input element and projects it beyond this hyperplane. This method is primarily based on the assumption that the model is completely linear. However, in most high-dimensional models, as in many deep neural networks, this is rarely the case. To overcome this problem, a linear approximation is first performed. The main problem with this attack is the inability to introduce domain-specific constraints and the significantly longer time required to generate adversarial instances compared to the FGSM.
	
	\paragraph{Jacobian-based Saliency Map Attack (JSMA)}
	JSMA is an iterative and targeted algorithm proposed by Nicolas Papernot \textit{et al. }\cite{papernot2016distillation} that uses a \textit{saliency map} to tell which feature has the greatest impact on classification. This saliency map is based on a \textit{jacobian matrix} which is a matrix containing the first-order partial derivatives as defined in Eq. \ref{eq:JSMA}
	\begin{equation}
		\label{eq:JSMA}
		J_F(x) = \frac{ \partial F(x)}{\partial x} = [\frac{\partial F_j(x)}{\partial x_i}] i \times j 
	\end{equation}
	This jacobian matrix, therefore, allows us to obtain the direction of sensitivity and, therefore know what input element influences the most desired output. This algorithm, based on the $L_0$ norm, has the advantage that it can generate adversarial samples using fewer features. It is therefore an interesting option for practical attacks against IDS.
	
	\paragraph{Carlini and Wagner (C\&W)}
	Carlini and Wagner \cite{carlini2017towards} proposed an optimization algorithm to generate adversarial examples under the $L_0$, $L_2$, and $L_\infty$ norms. This attack is different from L-BFGS because it uses a different loss function to escape box constraints. They redefine the initial problem of adversarial examples previously defined in Eq. \ref{eq:advex}.
	This redefinition is given in Eq. \ref{eq:carlini}
	\begin{equation}
		\label{eq:carlini}
		\begin{split}
			\text{Minimize: }D(x, x + \delta) + c.f(x + \delta) \\
			\text{Such that: }x + \delta \in [0,1]^n \text{----- constraint 2}
		\end{split}
	\end{equation}
	This attack is one of the most successful since it was able to break several defenses such as defensive distillation (see section \ref{Def_Dis}). This attack can be used in a targeted and non-targeted version. Nevertheless, even if C\&W is very efficient, it should also be noted that this algorithm takes significantly more time to generate adversarial instances.
	
	\paragraph{Elastic-Net Attacks to Deep Neural Networks (EAD)}
	This iterative algorithm proposed by Pin-Yu \textit{et al. }\cite{chen2018ead} introduces the use of the $L_1$ norm to generate perturbations to create adversarial examples. The authors transformed the problem into an elastic network regularized optimization problem. The elastic network regularization takes advantage of Lasso (using the $L_1$ norm) and Ridge (using the $L_2$ norm) regularization. EAD uses an iterative attack under $L_2$ using a $L_1$ regularizer. The original Elastic-Net regularization defined in Eq. \ref{eq:eadeq} is redefined for EAD as shown in Eq. \ref{eq:eadeq2}.
	
	\begin{equation}
		\label{eq:eadeq}
		\begin{split}
			\text{Minimize}_{z \in Z}\text{ }f(z) + \lambda_1 ||z||_1 + \lambda_2 ||z||^2_2 \\
		\end{split}
	\end{equation}
	\begin{equation}
		\label{eq:eadeq2}
		\begin{split}
			\text{Minimize}_x\text{ }c.f(x,t) + \beta ||x - x_0||_1 + ||x - x_0||^2_2 \\
			\text{Such that: }x \in [0,1]^n 
		\end{split}
	\end{equation}
	
	The experimental results \cite{chen2018ead} show that the attack is as effective as other state-of-the-art attacks. It is important to note that the results of this attack showed that it was the most effective in terms of transferability, which makes it interesting both for attackers using \textit{substitute models} (black-box attack) and also for defenders using the adversarial training defense. In addition, the authors showed that this attack, like C\&W, can break the defensive distillation defense. However, due to its optimization problem, it takes more time to execute than FGSM.
	
	\subsection{Black-Box algorithms}
	
	\paragraph{Zeroth-Order Optimization (ZOO)}
	It's a black-box and score-based algorithm inspired by the C\&W attack. As the name suggests, instead of using First-Order Optimization, it employs Zeroth-Order Optimization. It uses the logit values thanks to a zeroth order oracle to estimate the gradients. To estimate the gradients and Hessian, the authors \cite{chen2017zoo} use the symmetric quotient difference.
	
	To avoid detection by other defenses, Oracle queries must be reduced. To this end, ZOO employs three techniques, importance sampling, hierarchical attacks, and attack space reduction. The results of the experiments suggest that this attack is effective against ML models in black-box settings.
	
	It is important to note that while this technique has comparable performance to C\&W and yields a better attack success rate than substitute model attack, it is much more resource intensive than white-box algorithms, and even than the substitute model attack.
	
	\paragraph{Boundary}
	This iterative targeted/non-targeted decision-based attack created by Wieland Brendel \textit{et al. }\cite{brendel2017decision} is notably effective as it does not require gradient information and succeeds in defeating many existing defenses like defensive distillation and gradient masking defenses. Moreover, it is more realistic with respect to the other attacks as it doesn't rely on probabilities, but rather on the decision, which is what Machine Learning APIs typically provide. According to the authors, despite being a black-box attack, the Boundary attack produces a similar misclassification efficiency as other white-box attacks such as FGSM, C\&W, and DeepFool.
	
	Boundary uses a relatively simple and flexible algorithm. This attack uses a simple rejection sampling algorithm to track the decision boundary from the adversarial classification region to the non-adversarial region. The main drawback of this attack is that it uses an excessive number of iterations to find adversarial examples due to its brute-force nature. 
	
	\paragraph{OPT}
	The OPT attack, proposed by Minhao Cheng \textit{et al. }\cite{cheng2018query}, is an iterative decision-based black-box attack that can be targeted or untargeted. Being a decision-based attack means that it just needs the decisions rather than logits or probabilities. This optimization-based attack uses the Randomized Gradient-Free (RGF) method to estimate the gradient at each iteration rather than using the zeroth-order coordinate descent method, which provides lower performance. The RGF method is defined in Eq. \ref{eq:opteq} to estimate the gradient:
	\begin{equation}
		\label{eq:opteq}
		\hat{g} = \frac{g(\theta + \beta u) - g(\theta)}{\beta}.u
	\end{equation}
	where g is the search direction, g($\theta$) is the distance from x$_0$ to the nearest adversarial example along the direction $\theta$, $\beta$ > 0 is a parameter and u is a random Gaussian vector. 
	
	This attack uses the L$_2$ and L$_\infty$ norms to determine the perturbation rate to be applied and the binary search to evaluate the objective function. The results showed that the OPT attack was more efficient in terms of the number of queries required than the boundary attack, a similar attack in that it also uses only the model decision to be able to generate adversarial perturbations. In terms of performance, OPT was shown to be as efficient as many other state-of-the-art algorithms.
	
	Despite this, it requires performing a large number of queries, which can be detected by the victim's model if defense mechanisms are in place.
	
	\paragraph{Substitute Model attack}
	This method, used to perform adversarial attacks in a black box setting, was designed by Papernot \textit{et al. }\cite{papernot2016transferability}. It allows extracting the architecture of the model, the decision boundaries, and its functionalities. The goal is to try to mimic the original model using several queries to obtain $y = F^s(x)$, i.e., the given prediction of the original model must be equal to the prediction of the copied model. Once the model has similar behavior, state-of-the-art white-box attacks are used to generate adversarial examples. Using the transferability property, which is intrinsic to machine learning models, the original model can be fooled. This type of attack is less effective than white-box attacks but since it does not rely on gradient information, this attack is indeed feasible against non-differentiable models. In addition, it has been shown that the Substitution Model attack also defeats defenses based on gradient masking and defensive distillation.

	\paragraph{(Wasserstein) Generative Adversarial Network (GAN/WGAN)}
	GAN is an algorithmic architecture created by Goodfellow \textit{et al. }\cite{goodfellow2020generative} in 2014 and used to generate synthetic instances that resemble the original real instances. GAN uses two neural networks called \textit{discriminator} and \textit{generator}, both of which play a zero-sum adversarial game. The generator will try to create fake instances using a random normal distribution to fool the discriminator. The discriminator's goal is not to be fooled and to try to identify the false instances by learning from real instances contained in a dataset. As it goes along, the generator will try to learn to create more real instances. 
	
	WGAN \cite{arjovsky2017wasserstein} uses a different method than GAN to compute the probability distance between the two distributions. Instead of using the Jensen-Shannon divergence \cite{lin1991divergence}, which is based on the Kullback–Leibler divergence \cite{kullback1951information}, it uses the Earth-Mover's distance \cite{rubner2000earth}. The main advantages of WGAN are that it solves the vanishing gradient and mode collapse problems through more stable training. However, even if this mitigates the stability problem, the attack remains unpredictable and therefore unstable.
	
	\section{Defense strategies}
	\label{sec:defenses}
	The following defenses are designed to mitigate the effect of adversarial examples, such as those generated by the adversarial attacks discussed in Section \ref{sec:attacks}. One could divide defenses into two different types: \textbf{proactive}, where the idea is to prevent the model to be fooled by adversarial examples, and \textbf{reactive}, where the defense tries to detect adversarial examples during attacks. A list of the well-known defenses is in \autoref{tab:defenses}.
	
	\begin{table*}[]
		\centering
		\resizebox{\textwidth}{!}{%
			\def\arraystretch{2}
			\begin{tabular}{|c|c|c|c|c|c|}
				\hline
				\textbf{Defenses} &
				\textbf{Advantages} &
				\textbf{Disadvantages} &
				\textbf{Type of defense} &
				\textbf{Efficiency} &
				\begin{tabular}[c]{@{}c@{}}\textbf{Applicable}\\  \textbf{for IDS}\end{tabular} \\ \hline
				Adversarial Training &
				\begin{tabular}[c]{@{}c@{}}Effective against all attacks \\ Easy to implement\end{tabular} &
				\begin{tabular}[c]{@{}c@{}}Not effective against attacks different from those used \\ during the training phase and \\ there is a trade-off between robustness and accuracy\end{tabular} &
				Proactive &
				Could &
				Yes \\ \hline
				Adversarial Detection &
				Keeps the accuracy of the model &
				Some detection methods are proven ineffective &
				Reactive &
				Could &
				Yes \\ \hline
				Obfuscated Gradients &
				Effective against gradient-based attacks &
				\begin{tabular}[c]{@{}c@{}}Different method to bypass this defense \\ and not effective against transferability\end{tabular} &
				Proactive &
				/ &
				Yes \\ \hline
				Defensive Distillation &
				Distilled model is less sensitive to small perturbations &
				Shown to be ineffective &
				Proactive &
				/ &
				Yes \\ \hline
				Feature Squeezing &
				Good performance for image classification domain &
				Not suitable for tabular data &
				Proactive &
				/ &
				/ \\ \hline
				Ensemble Defense  &
				Combines several defense methods &
				Not efficient if the defenses used are broken &
				Reactive &
				/ &
				Could \\ \hline
				Feature Removal &
				Decreases the attack surface &
				Decreases the general performance of classification &
				Proactive &
				Yes &
				Yes \\ \hline
				Adversarial Query Detection &
				Don't modify the performance of the model &
				Applicable only for black-box attacks &
				Reactive &
				/ &
				Yes \\ \hline
			\end{tabular}%
		}
		\caption{List of well-known state-of-the-art evasion defenses}
		\label{tab:defenses}
	\end{table*}
	
	\subsection{Proactive defenses}
	
	\paragraph{Adversarial Training}
	The purpose of this defense proposed by Ian J. Goodfellow \textit{et al. }\cite{goodfellow2014explaining} is to strengthen the model against adversarial attacks by taking them into account during the learning phase. This defense can be seen in two ways. One can either provide the adversarial examples directly to the model with the training data or incorporate them into the loss function of the model which acts as a regularizer. This kind of defense is easy to implement and can be used very well in the IDS domain.
	
	A variant, called Ensemble Adversarial Training \cite{tramer2017ensemble} allows to improve robustness against attacks. The effectiveness of this defense can be improved by taking into account adversarial examples generated with different algorithms rather than using only one. If the model is trained with adversarial examples generated based on some adversarial attacks, an attacker could use other attacks to fool the classifier by exploiting the lack of generalization.
	
	Although the models become more robust to adversarial examples, they are not completely immune because some adversarial examples may go undetected. Moreover, this defense is limited by a trade-off between robustness and accuracy, as the more, the model is trained with adversarial examples, the more its overall performance decreases. Athalye \textit{et al. }\cite{athalye2018obfuscated} also showed that, if the model is trained with adversarial examples generated using the $L_\infty$ norm, the model is less robust as compared to training based on other norms ($L_0$, $L_1$ and $L_2$)
	
	\paragraph{Obfuscated Gradients}
	This technique is based on a gradient masking method so as to disrupt the descent of the gradient and, in this way, prevent gradient-based attacks from being able to successfully exploit the gradient by trying to make the model non-differentiable. 
	
	Gradient masking was broken by several attacks. One of them is in a white-box setting by using a random step and then switching to a gradient-based algorithm like FGSM for example. It was also broken in the black-box setting via transferability, which ensures the effectiveness of adversarial examples against other models than the one on which they were generated. Papernot \textit{et al. }\cite{papernot2017practical} show in particular that black box attacks are more effective than white box attacks when gradient masking defense is used. 	Furthermore, Athaye \textit{et al. }\cite{athalye2018obfuscated} have shown how to bypass three types of obfuscated gradients, namely: shattering gradients, stochastic gradients, and vanishing/exploding gradients.
	
	It can be noted that this defense could be used in the IDS domain but may not be as effective since, in theory, an attacker has limited knowledge of the defender's model, which limits the possibility of using white box attacks directly.

	\paragraph{Defensive Distillation}
	\label{Def_Dis}
	Initially used to reduce the dimensionality of DNN, a defense based on distillation is proposed by Papernot \textit{et al. }\cite{papernot2016distillation} defensive distillation aims to smooth the decision surface of the model. The distillation method uses two neural network models. An initial network, taking as input the training data and the corresponding labels. The model provides predictions in a probability vector and transfers this knowledge to the second network, called the distilled network. This network, therefore, takes the same training data as the initial network, but the corresponding labels are taken from the probability vector of the initial network, then new predictions are made. The authors showed that defensive distillation is less sensitive to small perturbations.
	
	This defense could be used in the IDS domain, however, Carlini and Wagner \cite{carlini2016defensive} have shown that this defense is broken by their attack. Therefore, it is not wise to use defenses that have already been shown to be broken when protecting a model.

	\paragraph{Feature Squeezing}
	This defense technique, proposed by Weilin Xu \textit{et al. }\cite{xu2017feature}, compresses the features of the instances and classifies them. Then a comparison is made between this classification and the classification of the original samples. If the results are different, then the instance is considered adversarial. Of the compression methods used by the author, such as bit depth compression, median smoothing, or non-local means, none consistently gives the best results, all need to be evaluated collectively as performance differs depending on the dataset used. 
	
	This defense is not suitable for the IDS use case because network traffic is often represented in tabular form and these compression techniques result in significant information loss for the underlying data.
	
	\paragraph{Ensemble Defense}
	This technique is expected to be effective by assuming that several different defense techniques improve the robustness of the model. Such a defense can be either proactive or reactive, or a mixture of both, in case the different defense mechanisms used are both reactive and proactive. It can therefore be used in the IDS domain and is potentially effective against various types of attacks.
	
	This technique is ineffective because an attacker can exploit any of the defenses' flaws to bypass them all. In addition, Warren He \textit{et al. }\cite{he2017adversarial} demonstrated that employing multiple weak defenses does not result in a stronger defense.
	
	\paragraph{Feature Removal}
	\label{def:FR}
	This defense consists of identifying the most vulnerable features and removing them from the data used to train the model. These vulnerabilities often come from the complexity of the model with high dimensions. This defense will therefore reduce the complexity of learning the model and remove some dimensions that are too vulnerable to escape the attack. The removal of features will reduce the attack surface by reducing the possible vectors that can be perturbed.
	
	However, the work of Apruzzese \textit{et al. }\cite{apruzzese2019evaluating} shows that feature removal decreases the performance of an IDS. It results in a loss of precision that increases the number of false positives.
	
	\subsection{Reactive defenses}
	
	\paragraph{Adversarial Detection} 
	
	This attack mechanism uses different methods to detect adversarial examples, based on statistical tools such as principal component analysis (PCA), distributions, and normalization. 
	
	One example of this method is the defense proposed by Feinman \textit{et al. }\cite{feinman2017detecting} which is based on two techniques: density estimates and Bayesian uncertainty estimates. The general idea of the first method is to check whether the density estimates of the last hidden layer for an input instance are significantly different from those associated with the training set containing the benign examples and, if so, the instance will be considered adversarial. Density estimates are made in the feature space of the last hidden layer because it is considered more linear than that of the input. Bayesian uncertainty estimates can be used to overcome situations where density estimates cannot detect adversarial examples. This method allows detection in low-confidence regions in the input space.
	
	The main advantage of this reactive defense is that it does not change the initial accuracy of the model. It could also be used in the IDS domain because it has no particular restrictions. However, it has two main problems: The first one is that it provides many false positives, which makes it less effective. The second problem is that most of these detection methods have been proven to be broken by Carlini and Wagner \cite{carlini2017adversarial}. Tianyu \textit{et al. }\cite{pang2018towards} proposed a more robust alternative to detect adversarial examples using kernel density estimates and the reverse cross-entropy training procedure.

	\paragraph{Adversarial Query Detection}
	Introduced in the Titi-taka framework proposed by Zhang \textit{et al. }\cite{zhang2020tiki}, this defense consists of detecting the number of abnormal queries that signal that an attack is being conducted. This defense reduces the number of possible queries sent to the model, making it more difficult to exploit for attacks using a large number of queries, while maintaining the initial accuracy. The main drawback is that this defense is only effective against black-box attacks that use large numbers of queries.

	\section{Adversarial attacks against IDS}
	\label{sec:studiedpaper}
	In recent years, many researchers have been interested in the implications of adversarial learning in the context of IDS by proposing several studies. We have selected several of them for their contributory aspect, and more precisely their contribution in terms of feasibility, a very important topic for IDS. These contributions are all listed in \autoref{tab:idsattack}. 
	
	\begin{table*}[]
		\centering
		\resizebox{0.9\textwidth}{!}{%
			\def\arraystretch{2}
			\begin{tabular}{|c|c|c|c|c|c|c|c|c|c|}
				\hline
				\textbf{Paper} &
				\textbf{Year} &
				\textbf{Adversarial attack} &
				\textbf{Target ML model} &
				\textbf{Datasets} &
				\textbf{Metrics} &
				\textbf{Defense used} &
				\begin{tabular}[c]{@{}c@{}}\textbf{Network} \\  \textbf{type}\end{tabular} &
				\begin{tabular}[c]{@{}c@{}}\textbf{Manipulation} \\  \textbf{space}\end{tabular} &
				\begin{tabular}[c]{@{}c@{}}\textbf{Domain} \\ \textbf{constraints}\end{tabular} \\ \hline
				\cite{rigaki2017adversarial} &
				2017 &
				\begin{tabular}[c]{@{}c@{}}FGSM\\ JSMA\end{tabular} &
				\begin{tabular}[c]{@{}c@{}}Decision Tree, Random Forest \\ Linear SVM, Voting Ensemble \\ DNN\end{tabular} &
				
				NSL-KDD &
				\begin{tabular}[c]{@{}c@{}}Accuracy \\ F1-score \\ AUC\end{tabular} &
				/ &
				Classical &
				Features &
				/ \\ \hline
				\cite{lin2022idsgan} &
				2018 &
				WGAN &
				\begin{tabular}[c]{@{}c@{}}Decision Tree, Random Forest \\ SVM, DNN, Naive Bayes\\ Logistic Regression, KNN\end{tabular} &
				
				NSL-KDD &
				Detection Rate &
				/ &
				Classical &
				Features &
				/ \\ \hline
				\cite{warzynski2018intrusion} &
				2018 &
				FGSM &
				DNN &
				
				NSL-KDD &
				\begin{tabular}[c]{@{}c@{}}Accuracy \\ FPR, FNR\\ Precision\end{tabular} &
				/ &
				Classical &
				Features &
				/ \\ \hline
				\cite{wang2018deep} &
				2018 &
				\begin{tabular}[c]{@{}c@{}}FGSM, JSMA\\ DeepFool\\ C\&W\end{tabular} &
				DNN &
				
				NSL-KDD &
				\begin{tabular}[c]{@{}c@{}}Accuracy, F1-score \\ FPR, Precision\\ Recall, AUC\end{tabular} &
				/ &
				Classical &
				Features &
				/ \\ \hline
				\cite{yang2018adversarial} &
				2018 &
				\begin{tabular}[c]{@{}c@{}}WGAN\\ ZOO\\ Substitute Model\end{tabular} &
				\begin{tabular}[c]{@{}c@{}}Random Forest \\ SVM \\ DNN\\ Naive Bayes\end{tabular} &
				
				NSL-KDD &
				\begin{tabular}[c]{@{}c@{}}Accuracy\\ F1-score, FPR\\ Precision, Recall\end{tabular} &
				/ &
				Classical &
				Features &
				/ \\ \hline
				\cite{apruzzese2018evading} &
				2018 &
				Manual perturbations &
				Random Forest &
				
				CTU-13 &
				\begin{tabular}[c]{@{}c@{}}Accuracy\\ F1-score\\ Precision, Recall\end{tabular} &
				/ &
				Classical &
				Features &
				/ \\ \hline
				\cite{martins2019analyzing} &
				2019 &
				\begin{tabular}[c]{@{}c@{}}FGSM, JSMA\\ DeepFool\\ C\&W\end{tabular} &
				\begin{tabular}[c]{@{}c@{}}Decision Tree, Random Forest\\ Naive Bayes, SVM, DNN\\ Denoising Autoencoder\end{tabular} &
				
				\begin{tabular}[c]{@{}c@{}}NSL-KDD\\ CICIDS-2017\end{tabular} &
				AUC &
				Adversarial Training &
				Classical &
				Features &
				/ \\ \hline
				\cite{clements2021rallying} &
				2019 &
				\begin{tabular}[c]{@{}c@{}}FGSM, JSMA\\ C\&W\\ EAD\end{tabular} &
				KitNET (Kitsune) &
				
				Kitsune &
				\begin{tabular}[c]{@{}c@{}}Success Rate\\ AUC\end{tabular} &
				/ &
				\begin{tabular}[c]{@{}c@{}}Classical\\ IoT\end{tabular} &
				Features &
				/ \\ \hline
				\cite{ibitoye2019analyzing} &
				2019 &
				\begin{tabular}[c]{@{}c@{}}FGSM\\ BIM\\ PGD\end{tabular} &
				DNN &
				
				BOT-IoT &
				Accuracy &
				Feature Normalization &
				IoT &
				Features &
				/ \\ \hline
				\cite{apruzzese2019evaluating} &
				2019 &
				\begin{tabular}[c]{@{}c@{}}Manual perturbations \\ modifying up to 4 features\end{tabular} &
				\begin{tabular}[c]{@{}c@{}}Decision Tree, Random Forest, SVM \\ DNN, Naive Bayes, Linear Regression, KNN\\ ExtraTrees, AdaBoostBagging, \\ Gradient Boosting, SGD Linear Classifier\end{tabular} &
				
				\begin{tabular}[c]{@{}c@{}}CTU-13\\ CICIDS-2017\\ CICIDS-2018\\ UNB-CA Botnet\end{tabular} &
				\begin{tabular}[c]{@{}c@{}}F1-score \\ Precision, Recall\\ Attack Severity\end{tabular} &
				Feature Removal &
				Classical &
				Features &
				/ \\ \hline
				\cite{hashemi2019towards} &
				2019 &
				\begin{tabular}[c]{@{}c@{}}Manual perturbations for packet-based IDS\\ A proposed optimization \\ attack for flow-based IDS\end{tabular} &
				\begin{tabular}[c]{@{}c@{}}KitNET (Kitsune)\\ DAGMM\\ BiGAN\end{tabular} &
				
				CICIDS-2017 &
				\begin{tabular}[c]{@{}c@{}}TPR\\ FPR\end{tabular} &
				/ &
				Classical &
				Traffic &
				Yes \\ \hline
				\cite{teuffenbach2020subverting} &
				2019 &
				\begin{tabular}[c]{@{}c@{}}Manual perturbations based on modification \\ of the payload size, packet\\  rate and bidirectional traffic\end{tabular} &
				\begin{tabular}[c]{@{}c@{}}Random Forest\\ SVM, KNN\\ Logistic Regression\end{tabular} &
				
				CICIDS-2017 &
				\begin{tabular}[c]{@{}c@{}}Accuracy, F1-score \\ Success Rate\\ TPR, FPR\end{tabular} &
				/ &
				Classical &
				Traffic &
				Yes \\ \hline
				\cite{sheatsley2022adversarial} &
				2020 &
				\begin{tabular}[c]{@{}c@{}}Adapted JSMA (AJSMA)\\ Histogram Sketch Generation (HSG)\end{tabular} &
				\begin{tabular}[c]{@{}c@{}}Decision Tree, SVM, DNN\\ Logistic Regression, KNN\end{tabular} &
				
				\begin{tabular}[c]{@{}c@{}}NSL-KDD\\ UNSW-NB15\end{tabular} &
				Success Rate &
				/ &
				Classical &
				Features &
				/ \\ \hline
				\cite{chen2020generating} &
				2020 &
				\begin{tabular}[c]{@{}c@{}}GAN\\ OPT\end{tabular} &
				\begin{tabular}[c]{@{}c@{}}Random Forest, OCSVM ,\\  DNN, Stacking Model \\ Naive Bayes\end{tabular} &
				
				\begin{tabular}[c]{@{}c@{}}Simulation of real ICS environment. \\ Real traffic captured \\ on the implemented infrastructure\end{tabular} &
				\begin{tabular}[c]{@{}c@{}}F1-score \\ Precision\\ Recall\\ Success Rate\end{tabular} &
				Adversarial Training &
				ICS &
				Traffic &
				Yes \\ \hline
				\cite{sadeghzadeh2021adversarial} &
				2020 &
				\begin{tabular}[c]{@{}c@{}}Adversarial Pad (AdvPad)\\ Adversarial Payload (AdvPay)\\ Adversarial Burst (AdvBurst)\end{tabular} &
				CNN &
				
				ISCXVPN2016 &
				\begin{tabular}[c]{@{}c@{}}F1-score \\ Precision\\ Recall\end{tabular} &
				/ &
				Classical &
				Traffic &
				Yes \\ \hline
				\cite{han2021evaluating} &
				2021 &
				\begin{tabular}[c]{@{}c@{}}Adversarial Features Generation with GAN \\ + Malicious Traffic Mutation with PSO\end{tabular} &
				\begin{tabular}[c]{@{}c@{}}Decision Tree, SVM\\ KitNET, DNN\\ Logistic Regression\\ Isolation Forest\end{tabular} &
				
				CICIDS-2017 &
				\begin{tabular}[c]{@{}c@{}}(evasive metrics)\\ MER, DER, PDR, MMR \\ (performance metrics)\\ F1-score, Precision, Recall\end{tabular} &
				\begin{tabular}[c]{@{}c@{}}Adversarial Training\\ Feature Removal\\ Adversarial Feature Reduction\end{tabular} &
				\begin{tabular}[c]{@{}c@{}}Classical\\ IoT\end{tabular} &
				Traffic &
				Yes \\ \hline
				\cite{zhang2022adversarial} &
				2022 &
				\begin{tabular}[c]{@{}c@{}}NES, Boundary, Pointwise\\ HopSkipJumpAttack\\ Opt-Attack\end{tabular} &
				\begin{tabular}[c]{@{}c@{}}MLP, CNN\\ C-LSTM\\ Stacking Model \end{tabular} &
				
				\begin{tabular}[c]{@{}c@{}}CICIDS-2017\\ CICIDS-2018\end{tabular} &
				\begin{tabular}[c]{@{}c@{}}Accuracy, F1-score \\ Precision, Recall\\ Success Rate\end{tabular} &
				\begin{tabular}[c]{@{}c@{}}Tiki-Taka\\  (Adversarial Training, \\ Adversarial Query Detection \\ and Ensemble Voting)\end{tabular} &
				Classical &
				Features &
				/ \\ \hline
				\cite{merzouk2022investigating} &
				2022 &
				\begin{tabular}[c]{@{}c@{}}FGSM, BIM\\ JSMA, DeepFool, C\&W\end{tabular} &
				MLP &
				
				NSL-KDD &
				\begin{tabular}[c]{@{}c@{}}Detection Rate\\ Lp norms mean and max\end{tabular} &
				/ &
				Classical &
				Features &
				/ \\ \hline
				\cite{debicha2022tad} &
				2022 &
				\begin{tabular}[c]{@{}c@{}}FGSM, PGD\\DeepFool, C\&W\end{tabular} &
				\begin{tabular}[c]{@{}c@{}} KitNET (Kitsune) \\ DNN  \end{tabular} &
				
				\begin{tabular}[c]{@{}c@{}} CICIDS-2017 \\ NSL-KDD \end{tabular}&
				\begin{tabular}[c]{@{}c@{}}Detection Rate\\ Precision, Recall, F-score\end{tabular} &
				\begin{tabular}[c]{@{}c@{}}Adversarial detection \\ based on Transfer learning\end{tabular} &
				Classical &
				Features &
				/ \\ \hline
			\end{tabular}%
		}
		\caption{List of recent evasion attacks against anomaly-based IDS}
		\label{tab:idsattack}
	\end{table*}

	In 2017, Maria Rigaki \textit{et al. }\cite{rigaki2017adversarial} demonstrated in their work that adversarial examples can be generated by adversarial algorithms to fool an IDS using a DNN trained on the NSL-KDD dataset. They showed that not all algorithms are suitable for fooling an IDS and pointed out the fact that FGSM is incompatible with this goal, but JSMA may be suitable. Furthermore, they showed that adversarial examples are capable of transferring to several machine learning models such as Decision Trees, Random Forests, Linear SVM, and Ensemble Voting. In addition, they showed that creating a feature-based adversarial instance requires knowing the mapping between features and network traffic and how the data is preprocessed because, unlike images, features extracted from network traffic are highly correlated.

	Zilong \textit{et al.} \cite{lin2022idsgan} proposed to study the effectiveness of adversarial examples with a WGAN using the NSL-KDD dataset. The performances of several classifiers were studied, namely decision tree, Random Forest, SVM, MLP, Naive Bayes, logistic regression, and KNN. The results showed that adversarial examples can fool all trained classifiers. They also showed that the attack remains effective even when using a limited feature space.

	Warzyński and Kołaczek \cite{warzynski2018intrusion} proposed a study on using FGSM to generate adversarial examples to fool a neural network-based classifier trained with the NSL-KDD dataset. The results showed that this attack is effective. It may be noted that this study does not address the feasibility of adversarial attacks in a realistic scenario and is limited to the study of a particular dataset and a particular attack.

	Zheng Wang \cite{wang2018deep} provides an in-depth analysis of the NSL-KDD dataset by investigating feature importance when generating adversarial examples against a multilayer perceptron (MLP) based IDS. He showed that the feasibility of adversarial attacks against IDSs is different from that of image classifiers by illustrating that not all adversarial algorithms are suitable for creating adversarial attacks against IDSs. Among these algorithms, JSMA seems to be the most suitable as it does not modify all features to create adversarial examples but only those it considers most important. This is particularly relevant since adversaries are usually limited in their ability to manipulate features due to restricted access or the complexity of manipulating them all at once. Feature importance shows that some features are more vulnerable than others in that they are more often selected by the algorithm during adversarial examples generation.
	
	Yang \textit{et al. }\cite{yang2018adversarial} proposed a more realistic approach to the problem by using three black-box attacks that assume no knowledge of the target model information. These three attacks are WGAN, ZOO, and Substitute Model. To analyze their performance, they took 5 classifiers, namely Random Forest, SVM, MLP, and Naive Bayes trained with the NSL-KDD dataset. The results showed that ZOO was the most efficient, the Substitute Model was the least efficient while WGAN provided good performance but was unstable due to its intrinsic properties. However, the paper does not discuss the feasibility of these attacks in real-world settings besides the black-box perspective.

	Instead of using state-of-the-art attacks, Apruzzese \textit{et al. }\cite{apruzzese2018evading} proposed an attack that iteratively produces manually defined perturbations. They studied the performance of this attack on a Random Forest classifier trained on the CTU-13 dataset which is based on a collection of Botnet attacks. This attack follows a fairly simple strategy, it clusters specific features and applies an iterative perturbation. The features are not chosen trivially, they are the ones that are easiest to manipulate, namely time, packet size, and the number of packets. The results showed that changing only a few features can lead to a decrease in classifier performance. From a feasibility point of view, this work is interesting because the modified features are at most four, and chosen in advance, which can be adjusted to modify only those features we have access to. 
	
	To generate adversarial examples, Martins \textit{et al. }\cite{martins2019analyzing} chose to work on the NSL-KDD dataset and a more realistic and recent dataset, namely CIC-IDS 2017. They also used adversarial training, first described in the image classification literature, to see if it could be applied to the IDS domain. The results showed that among the attacks used, JSMA is the least effective but disrupts the fewest features. They also showed that Adversarial Training improves the overall robustness of all classifiers, namely Decision Tree, Random Forest, Naive Bayes, SVM, DNN, and Denoising Autoencoder. Nevertheless, we can note that the feasibility of adversarial attacks has not been addressed in this work, except for the use of a more realistic dataset.

	Joseph \textit{et al. }\cite{clements2021rallying} used Kitsune's classifier called KitNET, and its dataset, to evaluate its robustness to adversarial examples generated based on four attacks (FGSM, JSMA, C\&W, EAD) using different norms ($L_0, L_1, L_2, L_\infty$). In a white box setting, the results showed that the classifier was vulnerable to all four attacks. 
	
	The study provided by Olakunle \textit{et al. }\cite{ibitoye2019analyzing} analyzed the impact of adversarial examples generated with FGSM, BIM, and PGD against two DNNs trained with the Bot-IoT dataset containing network attacks such as DOS or DDOS. The results showed that these adversarial attacks performed well in the IoT domain. In addition, they proposed feature normalization as a defense mechanism. The results showed that this defense was not effective as it increased the accuracy of the classifier however it also made the model more vulnerable to adversarial examples.
	
	According to Apruzzese \textit{et al. }\cite{apruzzese2019evaluating} on evaluating the effectiveness of adversarial attacks against botnet detectors, their results show that it is possible to fool this type of NIDS detector based on machine learning algorithms. They found that all the machine learning algorithms studied in this paper, namely Random Forest, Decision Trees, AdaBoost, Multi-Layer Perceptron, K-Nearest Neighbor, Gradient Boosting, Linear Regression, Support Vector Machines, Naive Bayes, ExtraTrees, Bagging, and Stochastic Gradient Descent Linear Classifier are susceptible to be fooled. In this study, the experiments are conducted from a more realistic perspective by taking into account some important domain constraints and assuming the gray box parameters, and using known realistic datasets containing botnet attacks to train their models. These datasets are as follows: CTU-13, IDS2017, CIC-IDS2018, and UNB-CA Botnet. To generate adversarial examples, the authors manually add small perturbations to a maximum of 4 features of each malicious instance keeping a realistic perspective. These features are duration, sent bytes, received bytes, and exchanged packets. Furthermore, they showed that using the defense called "\textit{feature removal}" (see Section \ref{def:FR}) does not guarantee robust protection for botnet detectors.
	
	Hashemi \textit{et al. }\cite{hashemi2019towards} propose a bottom-up approach by first analyzing the characteristics of the IDS datasets to understand their domain constraints. Once these constraints were identified, they showed that it is possible to fool different IDS models (Kitsune, DAGMM, and BiGAN) trained on the Kitsune and CICIDS2017 datasets, respectively, under these domain constraints for both packet-based and flow-based IDSs. To approach the problem more realistically, they proposed two algorithms for each of these network traffic types. For packets, the algorithm is divided into three parts: one function that generates delays between packets, another one that splits packets to have more packets, and the last one used to generate new packets. For flows, the algorithm uses a system of groups, only one of which can be modified and on which the attacker can apply perturbations. Recent work by Teuffenbach \textit{et al. }\cite{teuffenbach2020subverting} building on this work, also uses this grouping method to modify only relevant features. However, these groupings are slightly different from those proposed by Hashemi. Their results also showed that their method, involving domain constraints directly in their optimization problem, was effective in fooling the models (DNN, DBN, and AE) trained on CIC-IDS2017 and NSL-KDD.

	In their paper, Aiken \textit{et al. }\cite{aiken2019investigating} investigated the effectiveness of a novel adversarial example generation method focused on a SYN Flood DDoS attack. The results showed that the proposed algorithm is effective in fooling the classifiers (Random Forest, SVM, Logistic Regression, and KNN) trained on the SYN Flood attack present in the CIC-IDS 2017 dataset. The accuracy of the model fell to 0\% using the proposed algorithm. However, some characteristics of the instances are manipulated when they are not supposed to be since they are not easily modified in reality, such as the traffic from the victim. 
	
	Sheatsley \textit{et al. }\cite{sheatsley2022adversarial} study showed that, even when several NIDS-related domain constraints are considered, limiting the number of features that can be modified, it is possible to create realistic adversarial examples capable of fooling attack detectors using the AJSMA (Adapted JSMA) and HSG (Histogram Sketch Generation) adversarial algorithms. The experiments were conducted on the NSL-KDD and UNSW-NB15 datasets. They showed that domains with more restrictive constraints, such as NIDS, are no more robust than those with fewer constraints, such as image recognition. They also showed that these attacks were effective because of their transferability.
	
	Jiming Chen \textit{et al. }\cite{chen2020generating} studied the impact of adversarial examples on the domain of Industrial Control Systems (ICS). They took a more realistic approach by limiting their knowledge of the models used by the defender. They reproduced an ICS system to create a realistic environment and trained their MLP model directly on the extracted traffic. They then used two attack algorithms, GAN and OPT, to produce adversarial instances while taking into account domain constraints such as constraints related to the protocols used during the attacks. Their results showed that the ICS domain was also vulnerable to adversarial examples. In this study, several models were tested, namely, Random Forest, OCSVM (One-Class SVM), DNN, Stacking Model, and Naive Bayes. They proposed to use adversarial training and they found that this defense improved the robustness of the models studied. 
	
	In their paper, Sadeghzadeh \textit{et al. }\cite{sadeghzadeh2021adversarial} proposed a problem-based approach as opposed to a feature-based approach \cite{pierazzi2020intriguing}. The authors used three new attacks to manipulate network traffic called Adversarial Pad (AdvPad) which adds the perturbation to the packet, Adversarial Payload (AdvPay) which adds perturbation to the payload and Adversarial Burst (AdvBurst) which adds newly crafted packets. They propose to manipulate traffic concerning different types of services such as VoIP, mail protocols, file transfer, or P2P present in the ISCXVPN2016 dataset. The results showed that the classifier used (CNN) decreased its robustness due to the use of the three attacks.

	\begin{table*}[h]
		\centering
		\resizebox{\textwidth}{!}{%
			\def\arraystretch{2}
			\begin{tabular}{|c|c|c|c|}
				\hline
				\textbf{Dataset} &
				\textbf{Year of publication} &
				\textbf{Description} &
				\textbf{Reliability} \\ \hline
				NSL-KDD &
				2009 &
				\begin{tabular}[c]{@{}c@{}}Due to the problems related to attacks inside the KDDCUP'99 dataset,\\  this dataset has been proposed with more up-to-date attacks.\end{tabular} &
				\begin{tabular}[c]{@{}c@{}}It is old and not adapted to train classifiers today \\ because it contains old attacks not used today\end{tabular} \\ \hline
				CTU-13 &
				2013 &
				\begin{tabular}[c]{@{}c@{}}It's a dataset that contains different botnet and benign \\ network traffic provided by different datasets.\end{tabular} &
				\begin{tabular}[c]{@{}c@{}}This dataset, almost 10 years old, is less interesting \\ than some more recent and semantically correct ones.\end{tabular} \\ \hline
				UNB-CA Botnet &
				2014 &
				\begin{tabular}[c]{@{}c@{}}This dataset was developed to provide a focus on the realism,\\ generality, and semantics of botnet attacks.\end{tabular} &
				\begin{tabular}[c]{@{}c@{}}Since it is synthetically constructed and is an aggregation \\ of several datasets with Botnet attacks (real or not), it has some bias, \\ although it provides a good training ground given its many attack instances.\end{tabular} \\ \hline
				UNSW-NB15 &
				2015 &
				\begin{tabular}[c]{@{}c@{}}This synthetic dataset, created by the Cyber Range Lab of UNSW Canberra, provides \\ several types of network attacks such as DoS, Exploits, Reconnaissance or Backdoors.\\  Its diversity of attacks makes it an interesting data source.\end{tabular} &
				\begin{tabular}[c]{@{}c@{}}from a reliability point of view, it contains biases due to its synthetic profile\\  (as for most public datasets) and is relatively old,\\  thus less interesting than more recent ones with similar attacks.\end{tabular} \\ \hline
				ISCXVPN2016 &
				2016 &
				\begin{tabular}[c]{@{}c@{}}It's a realistic dataset that provides different types of traffic such as VoIP, \\ file transfers, P2P, and email protocol usage.\end{tabular} &
				\begin{tabular}[c]{@{}c@{}}It is reliable because all the traffic generated is real. \\ It also provides traffic from different services with the use of VPNs in some cases.\end{tabular} \\ \hline
				CIC-IDS2017 &
				2017 &
				\begin{tabular}[c]{@{}c@{}}Implement different network attacks like Bruteforce, DoS, Web, Botnet, Infiltration \\ and DDOS against machines in a simulated enterprise architecture.\end{tabular} &
				\begin{tabular}[c]{@{}c@{}}This dataset respect the 11 minimum requirements to have a reliable dataset \cite{gharib2016evaluation}. \\ Semantic logic is respected, and the list of attacks is relatively useful for classifiers today.\\ Some semantics flaws in attacks are solved in an update done by \cite{engelen2021troubleshooting} \end{tabular} \\ \hline
				CIC-IDS2018 &
				2018 &
				\begin{tabular}[c]{@{}c@{}}Implement different network attacks like Bruteforce, DoS, Web,\\  Botnet, Infiltration, and DDOS against machines in an AWS network.\end{tabular} &
				More recent than CIC-IDS2017 and follows the same requirements. \\ \hline
				Kitsune &
				2018 &
				\begin{tabular}[c]{@{}c@{}}The authors of Kitsune used a dataset specifically for the development of their IDS. \\ This dataset contains different types of attacks against a surveillance system. \\ These include reconnaissance attacks (OS Scan, Fuzzing), \\ MitM (ARP cache poisoning), DoS (SYN DoS), and Botnet (Mirai).\end{tabular} &
				\begin{tabular}[c]{@{}c@{}}This dataset is based on an implementation that mixes traffic\\  from a classical network and an IoT network.\\  The attacks are recent and therefore relevant.\end{tabular} \\ \hline
				BOT-IoT &
				2019 &
				\begin{tabular}[c]{@{}c@{}}Designed by the Cyber Range Lab of UNSW Canberra. \\ It contains different botnet attacks like DOS, DDOS, and Reconnaissance\end{tabular} &
				\begin{tabular}[c]{@{}c@{}}This dataset is more recent than previous ones and provide \\ good reliability in term updated attacks.\end{tabular} \\ \hline
			\end{tabular}%
		}
		\caption{List of the most popular datasets for training state-of-the-art IDS}
		\label{tab:datasets}
	\end{table*}
	
	Han \textit{et al. }\cite{han2021evaluating} proposed a novel attack using a GAN and a Particle Swarm Optimization (PSO) technique to directly manipulate the traffic under a black-box assumption. This approach is more realistic because the feature space is not easily reversible in the IDS domain, which means that the change in feature value cannot be transferred directly into the network traffic due to the numerous domain constraints. The attack process is divided into two parts, the GAN is used to generate adversarial features first, and then the PSO is used to add mutations to the malicious traffic. To evaluate the effectiveness of their attack, the authors used three new metrics to test the evasion effectiveness, namely the Detection Evasion Rate (DER), the Malicious Traffic Evasion Rate (MER), and the Probability Drop Rate (PDR) of the malicious traffic. They also proposed a new metric to provide an interpretability indicator called Malicious features Mimicry Rate (MMR) which provides a measure of how far the adversarial features are from the malicious features during mutation. The results showed that the attack was able to fool packet-based IDSs trained with the Kitsune dataset, as well as flow-based IDSs trained with the CIC-IDS2017 dataset.  However, the effectiveness of the attack varies depending on the knowledge of the extracted features. If the substitute model does not know any of the features extracted by the extractor, it will still be able to generate adversarial traffic capable of fooling the classifiers by randomly choosing less effective features. Despite the black-box assumption, the authors assumed that the extractor used by the IDS is known, which is not always the case in real settings.
	
	Zhang \textit{et al. }\cite{zhang2022adversarial} proposed a new version of their original paper \cite{zhang2020tiki}. Their goal was to evaluate their framework's performance in terms of improving IDS protection against evasion attacks subject to limited defense knowledge. In particular, they used multiple decision-based black-box algorithms to provide a more realistic representation of the problem. The results showed that the classifiers, using MLP, CNN, and C-LSTM, trained on the CIC-IDS2018 dataset were all vulnerable to the used black-box algorithms, namely NES, Boundary, Pointwise, HopSkipJumpAttack, and Opt-Attack. To improve the robustness of the classifiers, the authors propose to use their Tiki-Taka framework combining several defenses, Adversarial Training, Ensemble Voting, and Adversarial Query Detection. Thus, the authors combined two proactive defenses and a reactive one respectively. The results indicate that this combination of defenses is effective against the attacks studied on the CIC-IDS2017 and CIC-IDS-2018 datasets. 
	
	In their paper, Mohamed Amine Merzouk \textit{et al. }\cite{merzouk2022investigating} provide an in-depth analysis of the feasibility of state-of-the-art attacks against an IDS trained with NSL-KDD. The adversarial algorithms studied are FGSM, BIM, DeepFool, C\&W, and JSMA. They showed that these adversarial algorithms produce invalid Adversarial Examples (AEs) if applied directly without taking domain constraints. For example, some of the generated values were negative or out of bounds, exceeding their feasible limit. These AEs must meet certain criteria to be valid. In particular, they show four constraints, namely out-of-range values, non-binary values, membership in multiple categories, and semantic links.
	
	Debicha \textit{et al. }\cite{debicha2022tad} proposed an adversarial detector design based on transfer learning. They evaluated the effectiveness of using multiple strategically placed adversarial detectors versus a single adversarial detector for intrusion detection systems. Their experiments were conducted on two IDS architectures: a serial architecture and a Kitsune-inspired parallel architecture. They chose four evasion attacks to generate adversarial traffic, namely FGSM, PGD, DeepFool, and C\&W. Although the attacks are feature-based and not traffic-based, the author has taken into account the domain constraints to make them more feasible. Their defense is based on the implementation of multiple adversarial detectors, each receiving a subset of the information passed by the IDS and using a suitable fusion rule to combine their respective decisions. Using this defense, they were able to improve the detection rate over adversarial training.

	\section{Feasibility of the existing evasion attacks}
	\label{sec:discuss}
	After reviewing all these papers, we can see that most of them do not consider the realistic aspect, or only in a limited way regardless of the domain, whether in an IoT context or a traditional enterprise network. Here are the realistic aspects that should be taken into account for future studies of the impact of adversarial attacks on intrusion detection systems:
	
	First, it seems that most attacks based on feature-space manipulation do not provide a sufficiently realistic approach, as features cannot be easily transcribed back into network traffic once extracted and modified. In addition, semantic and syntactic constraints restrict the modifications applied to any feature. When working on the feature space, it is necessary to ensure that the generated adversarial network traffic is valid. An additional step must be performed after adding adversarial perturbations to the feature space. This step is called problem-space projection, which is the projection of the adversarial example into the realistic problem space. The objective of this step is to ensure that the reverse feature-mapping is doable as shown in \cite{pierazzi2020intriguing}.

	Second, the attacker is not supposed to know the mapping of raw network traffic into features, or at least not all of them, nor the semantic or syntactic links that exist between these features. This means that the assumptions about the feature extractor used in the model learning pipeline must remain limited if a truly realistic scenario is to be performed for the attacker. In other words, the attacker's knowledge should be limited and assumptions of full knowledge of the IDS should be avoided. 
	
	Third, some papers \cite{zhang2022adversarial,yang2018adversarial} have addressed the black box hypothesis using attacks such as Boundary, NES, OPT, or ZOO. In reality, these black box attacks are easily detected by simple defenses such as Query Detection. In addition to the explanation given in the previous two points, one cannot query the IDS like an oracle repeatedly because the attacker could easily reveal himself, in addition to the fact that IDSs are not designed to deliver feedback when queried.

	\section{Conclusion}
	\label{sec:conclu}
	Current research on the impact of using evasion attacks to bypass machine learning-based NIDS has shown that a slight perturbation can allow the attacker to circumvent detection. This of course raises a security concern, as the use of machine learning models is becoming more prevalent in the cybersecurity field. However, while it is theoretically possible to exploit these models, their exploitation is a bit different in a real-world setting. In this paper, we have reviewed the most recent attacks based on two possible adversarial strategies, namely the white-box and the black-box settings. We then explored some popular defense mechanisms. For each of the attacks and defenses, we elaborated on their suitability in the IDS domain. Finally, a set of related research papers are highlighted to clarify the feasibility of adversarial attacks in a more realistic context in the cybersecurity domain, specifically in the IDS domain. Concerning feasibility, we have provided several criticisms regarding recently published work by identifying their manipulation space. Thus, future research should focus on manipulating the traffic space by limiting the attacker's knowledge. We believe that this review has highlighted various points that may have been overlooked in some previous research, and that it will allow future research in this area to better address the various realistic constraints. 
	
	\bibliography{Mybiblio}
	
\end{document}